\documentclass[twoside,english,10pt,pra,showpacs,superscriptaddress,nobibnotes,aps,twocolumn]{revtex4-1}
\usepackage[T1]{fontenc}
\usepackage[latin9]{inputenc}
\usepackage{color}
\usepackage{babel}
\usepackage{amsmath}
\usepackage{amssymb}
\usepackage{graphicx}
\usepackage {hyperref}
\usepackage{breakurl}

\makeatletter

\providecommand{\tabularnewline}{\\}

\makeatother

\begin{document}

\preprint{This line only printed with preprint option}

\title{Fluctuation analysis of electric power loads in Europe: Correlation
multifractality vs. Distribution function multifractality}

\author{Hynek Lavi\v{c}ka}
\email{hynek.lavicka@fjfi.cvut.cz}

\thanks{Thanks to I. Jex, P. Exner, J. Tolar for support of the work.}

\affiliation{Alten Belgium N.V., Chaussée de Charleroi 112, B-1060 Bruxelles,
Belgium}

\affiliation{DYMO BVBA, Industriepark-Noord 30, 9100 Sint-Niklaas, Belgium}

\affiliation{Department of Institutional, Environmental and Experimental Economics,
University of Economics in Prague, W. Churchilla 4, CZ-130 67 Praha
3, Czech Republic}

\affiliation{Institute for Theoretical Physics, Celestijnenlaan 200D, KU Leuven,
B-3001 Leuven, Belgium}

\author{Ji\v{r}í Kracík}
\email{nyrlem.astro@seznam.cz}

\affiliation{Charles University in Prague, Faculty of Social Sciences, Institute
of Economic Studies, Opletalova 26. CZ-11000 Prague 1, Czech Republic}

\keywords{Electric power loads, MFDFA, Multifractality, Time series analysis,
Langevin equation, Machine learning}

\pacs{02.50.Fz, 05.10.Gg, 05.40.Fb, 05.45.Tp, 88.05.-b}
\begin{abstract}
We analyze the time series of the power loads of the 35 separated
countries publicly sharing hourly data through ENTSO-E platform for
more than 5 years. We apply the Multifractal Detrended Fluctuation
Analysis for the demonstration of the multifractal nature, autocorrelation
and the distribution function fundamentals. Additionally, we improved
the basic method described by Kanterhardt, et al using uniform shuffling
and surrogate the datasets to prove the robustness of the results
with respect to the non-linear effects of the processes. All the datasets
exhibit multifractality in the distribution function as well as in
the autocorrelation function. The basic differences between individual
states are manifested in the width of the multifractal spectra and
in the location of the maximum. We present the hypothesis about the
production portfolio and the export/import dependences.
\end{abstract}
\maketitle

\section{Introduction}

Recently unprecedented processes influenced the global economy; ``globalization''
as an integration of the global production system on one side and
a slow but stable banishment of the national overlook of the markets
- ``liberalization'' on the other side. The energy sector despite
of its crucial importance and appropriate level of the government
control was also influenced. In Europe, two joint markets, Nordpool
and CE, have been established and the production of the electric energy
is not necessarily produced and consumed within one country but  local
distributors are free to buy electric energy from abroad.

On the other hand, from the ecological point of view, the Kyoto protocol
and the Doha amendment prescribe the governments to focus on renewable
sources to assure transition to sustainable development. A general
feature of the renewable sources in contrast to the classical sources
is low yield as well as the fact that installations are placed far
from the centers of consumption that creates problems of additional
currents on the international scale, see e.g., \cite{key28}. We also
note that each country has specific portfolio of the power plants
and they operate to follow needs of  consumers. 

Electric grid is operated on national level by the Transmission System
Operator (TSO) that controls a swift and effective flow of energy
from the powerplants to the consumers. The process is based on the
large quantities of data measured in real-time and instantaneous action
is taken if necessary to balance the electric network to be proportional
to consumption.  Naturally, the process is influenced by deterministic
trends as well as random effects, see \cite{key02,Harris_Electricity_markets}.

In the framework of physics there was developed the Langevin equation
describing a path of a particle in a random environment

\begin{eqnarray}
\mathrm{d}X\left(t\right) & = & \mu\left(t,X\left(t\right)\right)\cdot\mathrm{d}t+\mathrm{d}W\left(t,X\left(t\right)\right),\label{eq:Langevin_equation}
\end{eqnarray}
where $\mu\left(t,X\left(t\right)\right)$ is the deterministic drift
that stands for deterministic effects in electric power system, while
$dW\left(t,X\left(t\right)\right)$ is a term representing noise depending
on actual state $X\left(t\right)$  that reflects random effects
within the electricity power grid. Analysis of Eq. \ref{eq:Langevin_equation}
in the framework of Brownian motion has been done in Ref. \cite{key05}.
Usual assumption on the properties of the noise is the Gaussian distribution
and memorylessness $\langle X\left(t\right)\cdot X\left(t+\Delta t\right)\rangle\sim\delta\left(\Delta t\right)$
or short range memory $\langle X\left(t\right)\cdot X\left(t+\Delta t\right)\rangle\sim\exp\left(-\frac{\Delta t}{L_{c}}\right)$,
where $L_{C}$ is a correlation length. 

However, in the real systems the properties may not be always satisfied,
e.g., the probability distribution is a non-Gaussian one. Even more,
it may be fat-tailed and (or) the autocorrelation function posses
long-range memory $\langle X\left(t\right)\cdot X\left(t+\Delta t\right)\rangle\sim\Delta t^{-\gamma}$,
where $\gamma$ is the exponent of the power law, see \cite{Rangarajan_Ding_Long-Range_correlations}.
A typical field of the study where the theory was used represents
a stock exchange, see \cite{key03}, where large datasets have been
stored and prepared for analysis. 

In recent decades  computers were widely used for the time-series
analysis and among the methods that are used belong the Detrended
Fluctuation Analysis (DFA) and its derivation the Multifractal Detrended
Fluctuation Analysis (MFDFA), see \cite{key06,key07,key23,key26,key27}.
Historically it is derived from the R/S method used in the field of
hydrology, see Ref. \cite{key29}. Recently the time series analysis
has focused on the problem of the trends and its impact on the analysis,
see Refs. \cite{key09,key36,key40}. The literature also contain a
non-orthodox modification of the MFDFA, see Ref. \cite{key35,key12}
as well as  modification usable for analyzing the cross-correlations
among the time series, see Ref. \cite{key40,key58}. We note that
the literature also contains modification that goes beyond one dimension,
see Ref. \cite{key60}.

Versatility of the method has been proved in a number of  applications.
Originally it was used to analyze the datasets in biophysics, see
Refs. \cite{key14,key15,key23} but recently the method is employed
in the number of the studies spanning from surface roughness via human
gain and econophysics to earthquakes and clouds, see \cite{key08,key03,key11,key13,key20,key25,key37,key61,key62,key64}.
Generally speaking, the method demonstrated extracts effectively the
information about the scaling properties within the dataset and then
the properties of the autocorrelation function.

Focusing on electroenergetics there is a number of the studies focusing
either on electric power loads and prices of electricity, see Ref.
\cite{key02,key16,key17,key41,key52}. Recently a competition of the
teams that predict price or electric power loads has grown  around
the globe and they use various methods, see Refs. \cite{key42,key45,key46}.
General weakness of number of studies is natural presence of oscillations
that spoils correct estimation of the Hurst exponent. The gap was
filled in \cite{kracik-lavicka:2016}.

In this study we would like to examine the datasets of the electric
power loads in the countries of Europe. We intend to perform the MFDFA
and we also plan to exploit various enhancements of the method to
reveal properties of both the distribution and autocorrelation functions.
Moreover we test how the properties of the functions are condensed
among parameters \textendash{} we intend to question multifractality
of the datasets. Finally we compare functionality of the states by
the multifractal properties.

The structure of the paper is as follows. Firstly, we describe the
method used in the study. Then we exploit power of  computers and
show the results of the analysis. Finally, we come with the conclusions
focusing on the stochastic properties of the datasets and then we
compare power loads of the European countries.

\section{Methods}

\begin{figure}
\includegraphics[scale=0.45]{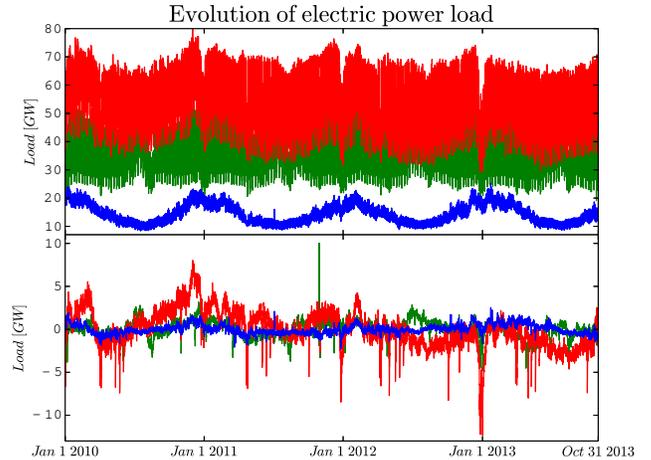}\caption{The electric power load of three countries of Europe (upper figure)
and the signals modulated $X^{car}\left(i\right)$ on the carrying
signal $X^{car}\left(i\right)$ after performing Step 1 (bottom figure)
spanning period since Jan 1 2010 to Oct 31 2013. Red, green and blue
curves are Germany, Italy and Norway respectively. We show the typical
examples of the the electric power load of countries with respect
to seasonality and inter-day and inter-week changes. Germany is an
example of the strong seasonality with the strong inter-day and inter-week
changes. Italy shows the strong inter-day and inter-week changes with
the limited seasonality. Finally, Norway exhibits the strong seasonality
with the limited inter-day and inter-week changes.}

\protect

\label{Power_load_in_Europe}
\end{figure}

\begin{figure}
\includegraphics[scale=0.43]{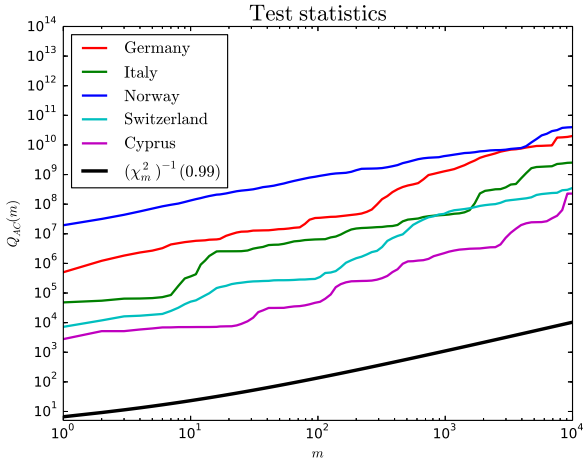}

\caption{The test statistics for the selected countries of Europe. Its comparison
with $\chi_{m}^{2}$-distribution at the $99\%$ significance level
with the $m$ degrees of freedom to the test hypothesis of none autocorrelations
(region $\left[0,\chi_{m}^{2}\left(0.99\right)\right]$).}

\protect

\label{Test_statistics}
\end{figure}

\begin{figure}
\includegraphics[scale=0.43]{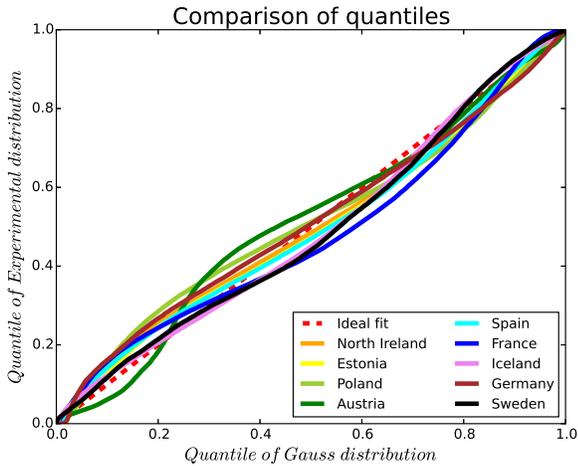}

\caption{Comparison of quantiles of stochastic part of the time series for
selected countries of the Europe.}

\protect

\label{Quantiles_plot}
\end{figure}

To analyze a timeserie $\left\{ X\left(i\right)\right\} {}_{i=1}^{N}$
with $N$ datapoints, usually  as a result of a measurement of a physical
variable, we perform modification of the Multifractal Detrended Fluctuation
Analysis (MFDFA). If the timeserie is non-stationary where the driving
forces of the system are stronger than the fluctuations within them
we propose a method that is based on the following 10 steps:

\paragraph{Fourier transform:}

Firstly we preform the Fourier transform to detect presence of the
carrier signal in the sample of the timeserie

\begin{eqnarray}
\widehat{X}\left(\omega\right) & = & \frac{1}{\sqrt{N}}\sum_{j=1}^{N}X\left(j\right)\exp\left(-2\pi\cdot\mathbf{i}\cdot j\cdot\omega\right).
\end{eqnarray}
Generalizations of the Fourier transforms are described in \cite{Kilbas_Saigo_H_transforms,Mathai_Hausbolt_Special_functions,Glaeske_Prudnikov_Skornik_Operational_calculus,Brichkov_Special_functions,Beals_Wong_Special_Functions,Davies_Integral_transforms,Erdelyi_Higher_Transcendental_Functions,Lebedev_Special_functions}
and they can be used to strengthen abilities of the method to the
wavelets of various types. We note that corresponding inversion formula
must be used below.

For the colored noise (fluctuations) the Fourier transform follows
$\hat{X}\left(\omega\right)\sim\vert\omega\vert^{\beta}$ with a number
of the additional peaks caused by the carrier signal upon the fluctuations
are modulated. We decompose the signal into two components

\begin{eqnarray}
\widehat{X}\left(\omega\right) & = & \widehat{X}^{fluc}\left(\omega\right)+\widehat{X}^{car}\left(\omega\right),
\end{eqnarray}

where $\widehat{X}^{fluc}\left(\omega\right)\sim\vert\omega\vert^{\beta}$
is the Fourier transform of the modulated signal and the Fourier transform
of the carrier signal is $\widehat{X}^{car}\left(\omega\right)$ and
it defines the profile average

\begin{eqnarray}
X^{car}\left(j\right) & = & \frac{1}{\sqrt{N}}\sum_{\omega=1}^{N}\widehat{X}^{car}\left(\omega\right)\cdot\exp\left(2\pi\cdot\mathbf{i}\cdot j\cdot\omega\right).
\end{eqnarray}

\paragraph{Test of presence of autocorrelations}

The first step of the procedure is to test of the presence of autocorrelation
within the sample. We define the test statistics made of the modulated
signal $X^{fluc}\left(i\right)$ as

\begin{equation}
Q_{AC}\left(m\right)=N^{2}\sum_{i=1}^{m}\frac{C_{i}^{2}}{N-i},
\end{equation}

where the sample autocorrelation is defined by formula

\begin{equation}
C_{i}=\frac{\sum_{k=i+1}^{N}X^{fluc}\left(k\right)\cdot X^{fluc}\left(k-i\right)}{\sum_{k=1}^{N}X^{fluc\ 2}\left(k\right)}.
\end{equation}

The test statistics $Q_{AC}\left(m\right)$ follows $\chi_{m}^{2}$
with $m$ degrees of freedom if none autocorrelations are present
in the sample. We fix the level of significance  $\alpha$ and varying
$m$. If $Q_{AC}\left(m\right)<\left(\chi_{m}^{2}\right)^{-1}\left(\alpha\right)$,
where $\left(\chi_{m}^{2}\right)^{-1}\left(\alpha\right)$ stands
for $\alpha$-th quantile of $\chi_{m}^{2}$ distribution \footnote{$\chi_{m}^{2}\left(\alpha\right)=\gamma\left(\frac{m}{2},\frac{\alpha}{2}\right)$,
where $\gamma$ is incomplete Gamma function.}, at the significance level $\alpha$ we conclude independence of
$x_{k}$s for the segment of length $m$. In the opposite case $Q_{AC}\left(m\right)>\left(\chi_{m}^{2}\right)^{-1}\left(\alpha\right)$
the hypothesis of none autocorrelations is rejected.

We admit that the step is optional and it serves as an indicator of
the presence of autocorrelations for the Gaussian distribution. The
indicator for the non-Gaussian distribution would need to derive appropriate
$\chi_{m}^{2}$ distribution.

\paragraph{Construction of profile}

Construction of a profile $X^{prof}\left(i\right)$ of the timeserie
from the modulated signal $X^{fluc}\left(i\right)$ is performed by
formula

\begin{equation}
X^{prof}\left(i\right)=\sum_{j=1}^{i}X^{fluc}\left(i\right).
\end{equation}

The step is auxiliary and it helps to perform the analysis accurately
even for the anti-persistent processes. Please, see Ref. \cite{key06}
for the discussion.

\paragraph{Separation into segments}

We separate the profile of the timeserie $X^{prof}\left(i\right)$
into the $N_{s}$ windows $X^{seg,w}\left(i\right)$ covering the
whole dataset with the length $s$ where each window is denoted by
a number $w$. The minimal number of the windows is $\lfloor\frac{N}{s}\rfloor$,
where $\lfloor x\rfloor$ is the largest integer smaller than or equal
to a number $x$. Thus the number of the timesteps $N$ is not generally
multiple of $s$. In order  to obtain better statistics we use as
small overlap of the consecutive windows where possible. 

\paragraph{Construction of trend}

We establish the local polynomial trend $X^{prof,w}\left(i\right)$
within each window $w$ of the size $s$ using the least square fit
of the dataset. Using the trend we detrend the data and we calculate
the sample variances in the window 
\begin{equation}
F^{2}\left(s,w\right)=\frac{1}{s}{\displaystyle {\textstyle \sum}}_{i=1}^{s}\left(X^{seg,w}\left(i\right)-X^{prof,w}\left(i\right)\right)^{2},
\end{equation}
where the calculation is performed for the windows $w=1,\ldots,N_{s}$.

\paragraph{Calculation of fluctuation function}

The fluctuation function of the $q$-th order is defined by the formula:

\begin{eqnarray}
F_{q}\left(s\right) & = & \begin{cases}
q\neq0 & \frac{1}{2N_{s}}\sum_{w=1}^{2N_{s}}\left(F^{2}\left(s,w\right)^{\frac{q}{2}}\right)^{\frac{1}{q}}\\
q=0 & \exp\left(\frac{1}{4N_{s}}\sum_{w=1}^{2N_{s}}\ln\left(F^{2}\left(s,w\right)\right)\right)
\end{cases}.\label{eq:fluctuation_function}
\end{eqnarray}

\paragraph{Calculation of generalized Hurst exponent}

The fluctuation function behaves as

\begin{eqnarray}
F_{q}\left(s\right) & \sim & s^{h\left(q\right)+1},\label{eq:generalized_hurst_exponent}
\end{eqnarray}
where $+1$ correction stands for the correction to the integrated
time series in $Step\ 2$. Generally, for positive $q$, $h\left(q\right)$
describes scaling behavior of large fluctuations within a segment
while for negative $q$, $h\left(q\right)$ describes scaling behavior
of small fluctuations. Independence $h\left(q\right)$ of $q$ means
monofractal behavior of the dataset. If we assume that the fluctuations
in the dataset are stable and follow the Gaussian distribution then
the Hurst exponent $H=h\left(q=2\right)$ is then $0<H<1$ for the
Gaussian distribution of the noise. For $H=\frac{1}{2}$, long range
autocorrelation is not present and the dataset may be independent.
 $H>\frac{1}{2}$ means long-range autocorrelations (persistence)
and $H<\frac{1}{2}$ stands for long range anti-autocorrelations (anti-persistence).

The non-Gaussian distributions with the power-law tail allows the
generalized Hurst exponent to exceed the range of $\left(0,1\right)$,
see \cite{key60} and $h\left(q\right)$ is a nontrivial function
of the variable $q$. It generally exhibits combination of the effects
caused by the probability distribution and the autocorrelation function.
In order to extract information on both we execute next two additional
steps.

\paragraph{Shuffling the dataset}

We shuffle the original timeserie $\left\{ x\left(i\right)\right\} {}_{i=1}^{N}$
to form $\left\{ x^{shuf}\left(i\right)\right\} _{i=1}^{N}$ using
the Fisher\textendash Yates algorithm, see \cite{Knuth_Volume_2}
for the description and the historical note. By calculating the fluctuation
function of the shuffled timeserie $F_{q}^{shuf}\left(s\right)$ performing
steps through $1$ to $5$ we obtain the shuffled fluctuation function

\begin{eqnarray}
F_{q}^{shuf}\left(s\right) & = & \overline{F_{q,\left\{ x^{shuf}\left(i\right)\right\} }\left(s\right)},\label{eq:shuffled_fluctuation_function}
\end{eqnarray}
where the averaging is executed for the different realizations of
shuffling. Shuffling of the original timeserie destroys (if present)
autocorrelations and thus the shuffled fluctuation function carries
information about the distribution function. By analogy of Eq. \ref{eq:generalized_hurst_exponent},
the shuffled fluctuation function behaves

\begin{eqnarray}
F_{q}^{shuf}\left(s\right) & \sim & s^{h^{shuf}\left(q\right)+1},\label{eq:shuffled_hurst_exponent}
\end{eqnarray}
where $h^{shuf}\left(q\right)$ called shuffled Hurst exponent describes
the scaling behavior of the fluctuation function of the shuffled timeserie.

\paragraph{Correlation Hurst exponent}

We define the autocorrelation Hurst exponent $h^{cor}\left(q\right)$
as follows:

\begin{equation}
h^{cor}\left(q\right)=h\left(q\right)-h^{shuf}\left(q\right).\label{eq:autocorrelation_hurst_exponent}
\end{equation}

The autocorrelation Hurst exponent contains information on the autocorrelation
function of the timeserie $\left\{ x\left(i\right)\right\} {}_{i=1}^{N}$.
 For $H^{cor}>0$ the timeserie is long range autocorrelated while
for $H^{cor}<0$ there is long range anti-autocorrelated behavior.
In case $H^{cor}=0$ the timeserie is either non-autocorrelated or
short range autocorrelated.

\paragraph{Surrogate dataset}

We perform additional test using the surrogate dataset. We intend
to verify robustness of our conclusions with respect to the non-linear
effects and the non-Gaussian features. The surrogate dataset conserves
the autocorrelation function. Calculation of the surrogate dataset
$\left\{ x^{sur}\left(i\right)\right\} {}_{i=1}^{N}$ is obtained
using the inverse Fourier transform of 
\begin{eqnarray}
\widehat{X}^{sur}\left(\omega\right) & = & \widehat{X}^{fluc}\left(\omega\right)\cdot\exp\left(\mathbf{i}\cdot Q\left(\omega\right)\right),
\end{eqnarray}
where $Q\left(\omega\right)$ is a IID random variable with uniform
distribution within the range $\left[-\pi,\pi\right]$. After then
the MFDFA is used on $\left\{ x^{sur}\left(i\right)\right\} _{i=1}^{N}$
using the above steps from 4 to 8. We perform $50$ samples of the
surrogate dataset and the surrogate fluctuation function is defined
by formula

\begin{eqnarray}
F_{q}^{sur}\left(s\right) & = & \overline{F_{q,\left\{ x_{i}^{sur}\right\} }\left(s\right)}.\label{eq:surrogate_fluctuation_function}
\end{eqnarray}

Analogically, the surrogate fluctuation function follows 
\begin{eqnarray}
F_{q}^{sur}\left(s\right) & \sim & s^{h^{sur}\left(q\right)+1},\label{eq:surrogate_hurst_exponent}
\end{eqnarray}
 where $h^{sur}\left(q\right)$ is the surrogate Hurst exponent.

\paragraph{Distribution Hurst exponent}

The distribution Hurst exponent is defined by 

\begin{equation}
h^{dist}\left(q\right)=h\left(q\right)-h^{sur}\left(q\right).\label{eq:distribution_hurst_exponent}
\end{equation}

The distribution Hurst exponent contains information about the distribution
function and its scaling properties.

\paragraph{Multifractal spectrum }

The fundamental properties of the autocorrelation function and the
distribution function of the time series can be studied using multifractal
spectrum $f\left(\alpha\right)$ that is Legendre transform of scaling
function $\tau\left(q\right)=q\cdot h\left(q\right)-1$

\begin{eqnarray}
f\left(\pi\right) & \equiv & q\left(\pi\right)\cdot\pi-\tau\left(q\left(\pi\right)\right)\label{eq:Legendre_transform}\\
\pi & \equiv & \frac{d\tau}{dq}=q\cdot h'\left(q\right)+h\left(q\right).\label{eq:Canonical_momentum}
\end{eqnarray}

Since we deal with three generalized Hurst exponents ($h\left(q\right)$,
$h^{cor}\left(q\right)$, $h^{shuf}\left(q\right)$) we calculate
$f\left(\pi\right)$, $f^{cor}\left(\pi\right)$ and $f^{shuf}\left(\pi\right)$.
In case that the autocorrelation function has monofractal property
or the distribution function has single exponent then $\pi=const.$
For multifractal case there occurs a distribution of $\alpha$ values
for both the autocorrelations as well as distributions, see \cite{Harte:2001}.
The width of the spectrum $\triangle\pi=\pi_{max}-\pi_{min}$ describes
the strength of multifractality, where $\pi_{max}=\max_{q}\pi\left(q\right)$
and $\pi_{min}=\min_{q}\pi\left(q\right)$. For broad spectrum $\triangle\pi$
is an indicator of strong multifractality while for narrow spectrum
it is the indicator of weak multifractality of the time series.

Analogically the autocorrelation and distribution multifractal spectrums
are defined by equations \ref{eq:Legendre_transform} and \ref{eq:Canonical_momentum}
with $h^{cor}\left(q\right)$ and $h^{dist}\left(q\right)$ respectively.
They produce $f^{cor}\left(\pi\right)$ and $f^{dist}\left(\pi\right)$
and the widths of multifractal spectrums are $\Delta\pi^{cor}=\pi_{max}^{cor}-\pi_{min}^{cor}$
and $\Delta\pi^{dist}=\pi_{max}^{dist}-\pi_{min}^{dist}$.

\begin{figure}
\includegraphics[scale=0.45]{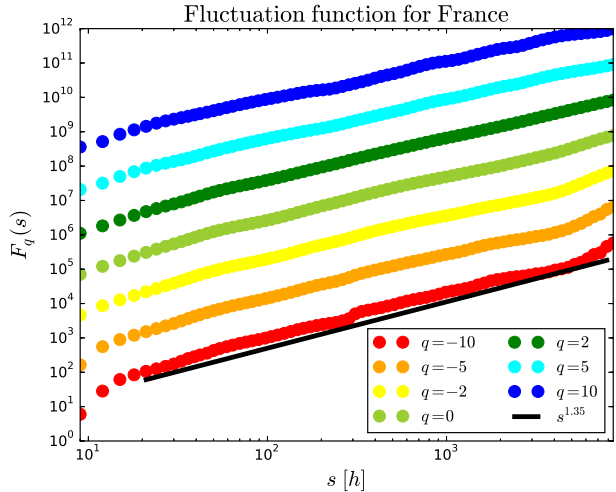}

\caption{The fluctuation function for the MFDFA-4 for France for $q\in\left\{ -10,-5,-2,2,5,10\right\} $.
Fluctuation functions are shifted by power of $10$ between each consecutive
$q$. The typical power law behavior is presented.}

\protect\label{Fluctuation function}
\end{figure}

\begin{figure}
\includegraphics[scale=0.45]{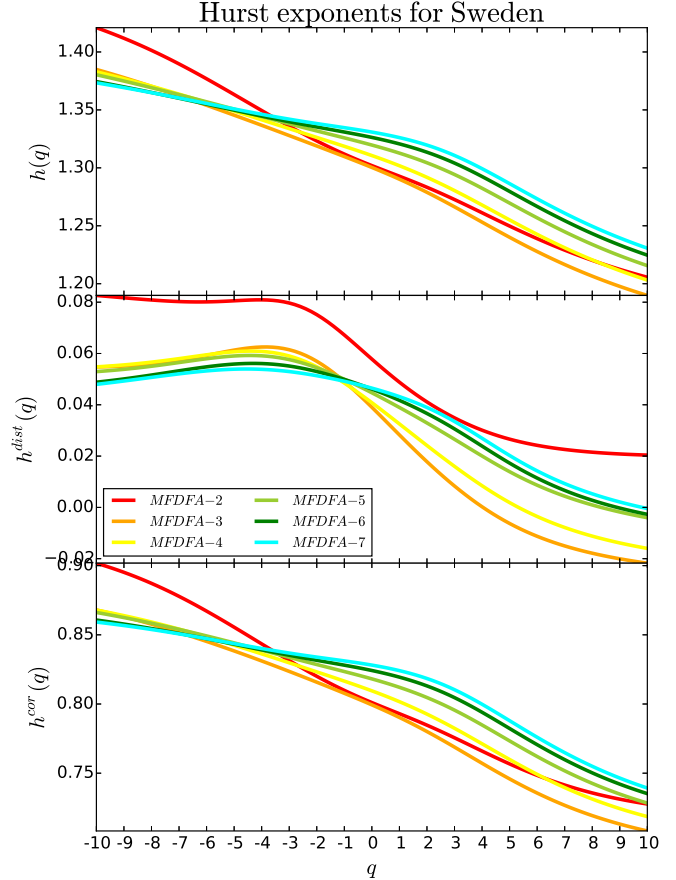}

\caption{The generalized Hurst exponents $h\left(q\right)$, $h^{shuf}\left(q\right)$
and $h^{cor}\left(q\right)$ are shown from top to bottom respectively.
The orders of the MFDFA between $2$ to $6$ are plotted with various
colors for United Kingdom (excluding Northern Ireland).}

\protect\label{Hurst exponent}
\end{figure}

\begin{figure}
\includegraphics[scale=0.45]{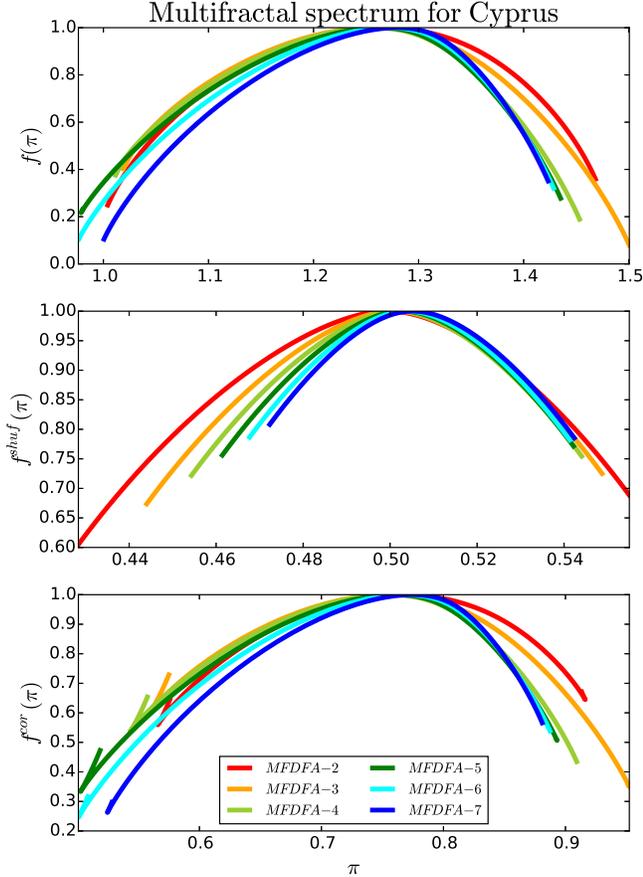}

\caption{The multifractal spectrum $f\left(\pi\right)$, the shuffled multifractal
spectrum $f^{shuf}\left(\pi\right)$ and the multifractal autocorrelation
spectrum $f^{cor}\left(\pi\right)$ for Cyprus of the MFDFA orders
between 2 and 7 is shown. The typical numerical issues can be observed
at the bottom as the wings.}

\protect\label{fig:multifractal spectra}
\end{figure}

\begin{figure}
\includegraphics[scale=0.45]{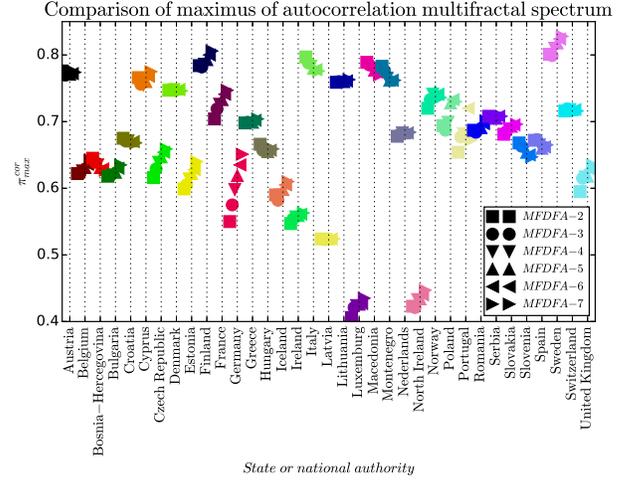}

\caption{Plot of maxims of the autocorrelation multifractal spectrum for states
or national authorities in Europe shown for various orders of the
MFDFA from 2 to 7.}

\protect\label{fig:mamimums_of_correlation_multifractal_spectrum}
\end{figure}

\begin{figure}
\includegraphics[scale=0.45]{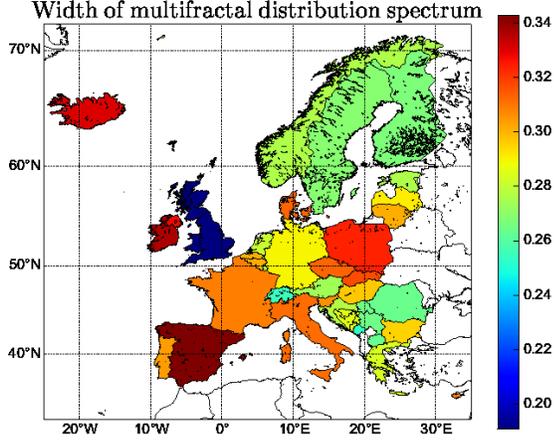}

\includegraphics[scale=0.45]{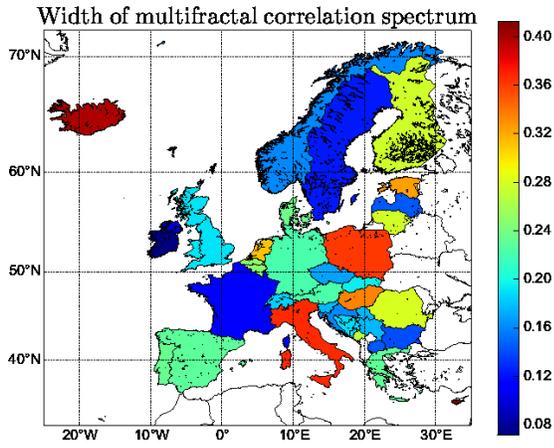}

\caption{The width of the multifractal shuffled (on the top) and correlation
(at the bottom) spectrum calculated by the MFDFA-4 and it is visualized
on the map of Europe. }

\protect\label{map_of_Europe}
\end{figure}

\begin{figure}
\includegraphics[scale=0.45]{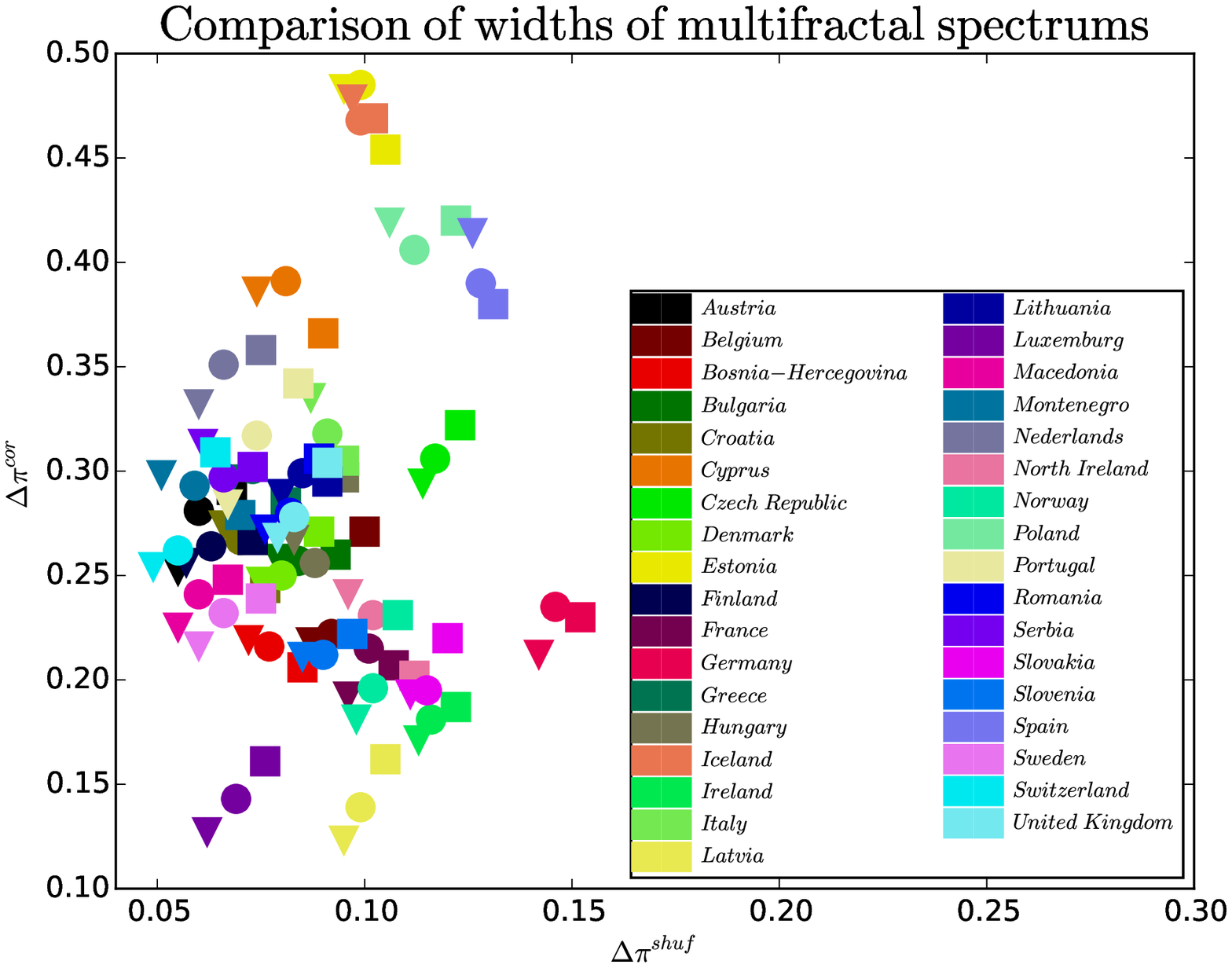}

\caption{Comparison of the widths of the correlated and shuffled multifractal
spectrums. The order of the MFDFA is indicated by different symbol.
Square, circle and triable down are the orders 4,5 and 6 respectively.}

\protect\label{fig:comparison_of_widths}
\end{figure}

\section{Results of analysis}

We perform the methodology which was described in the previous section
on a dataset of the electric power load for 35 European countries
or the independent electric power systems which were obtained from
ENTSO-E\footnote{Web page \url{http://www.entsoe.eu}} . This organization
 roofs local Transmission System Operators (TSO) that control the
swift and effective functionality of the backbone electric power grid
in the European countries. The dataset spans from January 1 2008 to
December 31st 2012 with the one-hour resolution for the longest dataset.

Each measured datapoint collects produced electricity within a region
as it counts for exports, imports and adding balance of pumped-storage
hydroplants \footnote{The datasets does not include electric power production for local
needs \textendash{} powerplants for a factory or a house \textendash{}
and also power production within islands disconnected from the mainland.}. It amounts electric power that is provided to users \textendash{}
citizens, industry and services \textendash{} through local electric
distribution companies. We also note that actual demand that is intended
to be satisfied must be lower in longer periods of time than actual
supply, otherwise blackout can happen on small or on the larger scale.
To prevent significant economic and physical waste of electric power
companies producing and transmitting electric power are trying to
 correlate both variables to the extent possible. We may also perceive
the dataset as a measure of an activity (economic, social) of a society.
Autocorrelations of the datasets can be understood as an indicator
of autocorrelations in European countries that share data through
ENTSO-E.

\subsection{Oscillations in power grid data}

Firstly, we show examples of datasets of typical European countries
at the top of the Fig. \ref{Power_load_in_Europe}. Each dataset exhibits
oscillations of period a year, a week and a day but the countries
differ in proportions of the oscillations on the time scale. Countries
of the south Europe usually exhibit limited one-year oscillations
in contrast to the countries of the northern Europe. Since most of
the countries are of Christian origin,  the religious holidays are
usually public holidays, e.g., Christmas and Easter. There is a significant
decrease of the electric power load during that period. Oscillations
can generally spoil the results of MF-DFA and biasing statistical
analysis, see \cite{key09,key36}. Performing Step 1 of the method
we obtain a dataset (signal) that is modulated on the carrying signal.
We note that performing cuts during Step 1 have been tested for different
exponents $\beta$ with minimal qualitative impact on results of following
analysis, see \cite{kracik-lavicka:2016}. 

We show examples of the modulated signal for Germany, Italy and Norway
at the bottom of the Fig. \ref{Power_load_in_Europe}. The modulated
signal $\left\{ X^{fluc}\left(i\right)\right\} _{i=1}^{N}$ is then
analyzed and tested for the presence of autocorrelations or  types
of distribution. We performed Komogorov-Smirnov test of the signal
and rejected null hypothesis that there is the Gaussian distribution,
e.g. for Norway the p-value stands at extremely small value $\thicksim10^{-290}$
and for Germany at $\thicksim10^{-105}$ so that it fails to reject
the null hypothesis. To demonstrate deviations from the Gaussian distribution,
see Fig. \ref{Quantiles_plot}. The power loads of the countries deviate
from diagonal line and the deviations show the presence of the asymmetric
probability distribution. Due to the fact that Austria does not fit
the pattern of the rest of the states we suppose that it is due to
diverge portfolio of the power plants.

Secondly we perform an autocorrelation test of a dataset, calculating
$Q_{AC}\left(m\right)$ and comparing with $\chi_{m}^{2}$ distribution
at $99\%$ level of significance. In the Fig. \ref{Test_statistics},
the test rejects hypothesis of no-autocorrelation within a sample
dataset for each European country. It motivates taking the following
steps of the methodology that aims at the calculation of the MF-DFA
to reveal properties of the autocorrelation and distribution function.
Via performing steps from 4 to 8 and 10 we obtain $F_{q}\left(s\right)$,
$F_{q}^{shuf}\left(s\right)$ and $F_{q}^{sur}\left(s\right)$ that
follows the power laws \ref{eq:generalized_hurst_exponent}, \ref{eq:shuffled_hurst_exponent}
and \ref{eq:surrogate_hurst_exponent}, which is for France demonstrated
in the Fig. \ref{Fluctuation function}.

\subsection{MFDFA}

The calculation of the appropriate Hurst exponents $h\left(q\right)$,
$h^{shuf}\left(q\right)$ and $h^{sur}\left(q\right)$ is then a key
point for calculation $h^{cor}\left(q\right)$ and $h^{dist}\left(q\right)$
(steps 9 and 11) that contain information on the scaling properties
of the autocorrelation and distribution function.  For illustration,
we show $h\left(q\right)$, $h^{dist}\left(q\right)$ and $h^{cor}\left(q\right)$
in the Fig. \ref{Hurst exponent} for Sweden. We note that the generalized
Hurst exponent $h\left(q\right)$ is estimated above standard range
for the Gaussian distribution $\left[0,1\right]$ \footnote{Then the autocorrelation Hurst exponent defined by Eq. \ref{eq:autocorrelation_hurst_exponent}
is in range $\left[-\frac{1}{2},\frac{1}{2}\right]$.}. Next $h^{dist}\left(q\right)$ is close to $0$ indicating that
the non-linear effects does not spoil Hurst exponent estimation. Finally,
the autocorrelation Hurst exponent $h^{cor}\left(q\right)$ indicates
a presence of multifractality for autocorrelation function.

Next, we focus on multifractal spectrums defined by Eqs. \ref{eq:Legendre_transform}
and \ref{eq:Canonical_momentum}, the multifractal spectrum $f\left(\pi\right)$,
the shuffled multifractal spectrum $f^{shuf}\left(\pi\right)$ and
the autocorrelation multifractal spectrum $f^{cor}\left(\pi\right)$
that are shown in Fig. \ref{fig:multifractal spectra} for Cyprus.
All spectrums exhibit a peak-like structure with the most frequent
value $\pi_{max}$, $\pi_{max}^{shuf}$ and $\pi_{max}^{cor}$ respectively,
forming the top of the appropriate multifractal spectrum. In table
\ref{multifractal_width_table} we show $\pi_{max}^{shuf}$ for the
states of Europe and it is mostly close to $\frac{1}{2}$, indicating
the most frequent presence of the Gaussian distribution. However,
maximums of the autocorrelation multifractal spectrum $\pi_{max}^{cor}$
varyes significantly from one country to another, see the Fig. \ref{fig:mamimums_of_correlation_multifractal_spectrum}.
$\pi_{max}^{cor}$ is also out of standard range $\left[-\frac{1}{2},\frac{1}{2}\right]$
for the Brownian processes except for North Ireland, Luxembourg and
Latvia \footnote{Well, Latvia is just at the border.} that is on
the edge. Most frequently the process is governed by \ref{eq:Langevin_equation}
and the stochastic process is persistent. However, the other states
are governed by the most frequent process described by equation 
\begin{eqnarray}
\mathrm{d}\dot{X}\left(t\right) & = & \mu\left(t,\dot{X}\left(t\right)\right)\cdot\mathrm{d}t+\mathrm{d}W\left(t,\dot{X}\left(t\right)\right).\label{eq:generalized_langevin_equation}
\end{eqnarray}
The maximum of the autocorrelation multifractal spectrum is then deceased
by 1 (Similarly when we skip to execute step 2 but execute steps 7,8
and 10 as they are). The process is then anti-persistent, similar
analysis in Ref. \cite{key41} also revealed anti-persistent processes
within the power grids. While the$2$nd order differential equations
are more stable with respect to the observation of displacement of
particles at the start and at the end than the $1$st order equations
and the weaker anti-persistence also decreases deviations. We argue
that higher $\pi_{max}^{cor}$ is more beneficial for stability of
the electric power grids.

\subsection{Multifractal spectra}

We also report that the wide multifractal spectrums are present within
the dataset, see Tab. \ref{multifractal_width_table}.  Since the
width of the autocorrelation multifractal spectrum $\triangle\pi^{cor}$
is wider, we conclude that multifractality of autocorrelation function
is also stronger than the distribution function. Since the width
of the distribution multifractal spectrum $\bigtriangleup\pi^{dist}$
is far smaller than $\triangle\pi^{cor}$ and $\triangle\pi$ and
$\pi_{max}^{dist}$ is close to zero, we conclude that the non-linear
effects are limited. Nevertheless they are present and multifractality
is not a side effect but it has a systematic nature in the operation
of the electric power system. We admit that the width of the distribution
and correlation spectrums is usually influenced by the numerical issues,
see the bottom of the Fig. \ref{fig:multifractal spectra}. 

\subsubsection{Operational properties of power grids in Europe}

Focusing on the operational properties of the national power grids,
we demonstrate the differences in the Fig. \ref{map_of_Europe} by
metric of the width of the shuffled and autocorrelation multifractal
spectrums. North-South or East-West spatial clustering based on political
or climatic is not present for both spectrums. The widest widths of
the shuffled multifractal spectrums is present for North Ireland,
Spain and a cluster of states in the central Europe, namely Germany,
Poland, Czech Republic and Slovakia. Interestingly, the cluster is
investigated in Ref. \cite{key28} from the point of view of the trans-border
flows reveals enormous production of the electric energy from the
wind power plants in north Germany and consumption in south Germany
with the enormous cross-border electric power traffic spanning Poland,
Czech Republic and Slovakia. 

The autocorrelation multifractal spectrum is usually wider than the
shuffled one. We observe central block of countries with relatively
narrow spectrum where countries with wider spectrum are placed randomly.
The widest spectrum is present for Iceland, Poland, Estonia and Spain
while the narrowest spectrum is present for France, Latvia and Bulgaria.
We refer reader to the Table \ref{multifractal_width_table} where
results of MFDFA for order 4 are presented. 

We show the space of the widths of the shuffled and autocorrelation
multifractal spectrums in the Fig. \ref{fig:comparison_of_widths}.
Most states operate within an elliptic cluster with center at $\left(0.2,0.2\right)$
a semi-axes $\left(0.05,0.1\right)$ with few deviations like Iceland,
Latvia, Spain and Poland which operate with much wider spectrums.
Additionally North Ireland, Latvia, Luxembourg and to some degree
additional states develop, due to their location of maximums of the
autocorrelation spectrum $\pi_{max}^{cor}$ close to $\frac{1}{2}$
and considerable widths of spectrums,  Formulas \ref{eq:Langevin_equation}
and \ref{eq:generalized_langevin_equation}. 

\subsection{Clusterization of the states}

\begin{figure}
\includegraphics[scale=0.4]{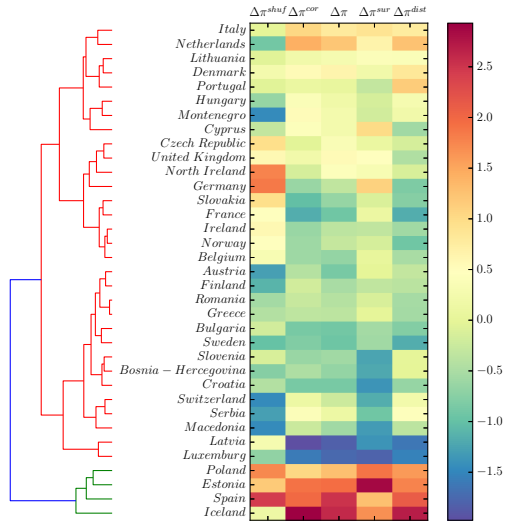}

\caption{The renormalized values of various widths of the multifractal spectra
are shown and compared for order 5 of the method. On left hand side
we present clusterization of the states in euclidean metric.}

\label{fig:Clusterization_of_states}
\end{figure}

We employed Machine learning tools to search for clusters to put out
previous observations of clusterization on solid ground, see \cite{Python_data_science_essentials,pedregosa2011scikit,Learning_scikit_learn,Introduction_Machine_Learning}.
Since the widths of multifractal spectra are order dependent the clusters
presence vary and more than 2 clusters were observed for orders 2,
3 and 4 for different methods of clusterization, metric and subset
of the widths of spectra.

Most countries of Europe operate  power grid systems in the region
with a limited variety of Hurst exponents, see red cluster in Fig.
\ref{fig:Clusterization_of_states}. The deviations are present and
that are namely Iceland, Spain, Estonia and Poland(only for the order
5). However, we observed additional presence of Germany, Czech Republic,
Slovakia, Portugal and North Ireland for different setup of clusterization
algorithm. For Iceland, portfolio of sources of electricity is widely
different than in the rest of Europe. Presence of Germany, Czech Republic,
Slovakia a Poland in deviating group may be caused by crossborder
transfer in cetral Europe, see \cite{key28}.

\subsection{Note on implementation and robustness of results}

We note that we performed the analysis for various orders of the method
with minimal impact on the conclusions. In the table \ref{multifractal_width_table}
we report the complete multifractal analysis of European countries
including the widths of all multifractal spectrums as well as the
locations of the maxims of all multifractal spectrums for order 4
of the MFDFA. 

The method have been implemented in \emph{Zarja} library\footnote{http://sourceforge.net/projects/zarja/}
with appropriate Python interface. Python is then used for post-processing
of the datasets. During the analysis we employed Matplotlib \cite{Hunter:2007},
Basemap, SciPy \cite{Scipy}, NumPy and Scikit tools \cite{pedregosa2011scikit,Learning_scikit_learn}.
The analysis using the MFDFA gives the best results for orders 4,
5 and 6. The lower orders lacks effective detrending but electic power
grids are relatively stable in this respect. 

On the other hand, we also admit that the analysis can be inaccurate
for both large $\vert q\vert$ and the high orders of the MFDFA. It
is caused by use of floating point numbers (IEEE 754) that  are inaccurate,
see \cite{Knuth_Volume_2}, due to the fixed length of the data-type.
To solve the problem we refer to the use the fractions or to the variable
length numbers, however the use is at the expense of the speed. Implementing
MFDFA can solve the problem.

\section{Conclusions}

We performed the modification of the MFDFA method on the dataset obtained
from the electric power grid of the European countries which consists
of two level of detrending. Dataset  analyzed are the electric power
loads with one hour frequency. In the study we focus on the properties
of the autocorrelation function and the probability distribution of
the stochastic process governing evolution of the system. 

To address whether the properties of the distribution and autocorrelation
function are unique or more parameters are involved we employ multifractal
spectrum of the autocorrelation and distribution function. All countries
admit multifractality of both the autocorrelation and distribution
functions. To be more specific, all datasets admit existence of the
non-Gaussian probability distribution due to the width of multifractal
spectra as well as the long-range autocorrelation function. 

We detected presence of the countries that posses the parameters that
Langevin equation for both the coordinates and the velocities have
to be employed. Generally, we verified the robustness of our analysis
performing the surrogated datasets as well as various orders of detrending
and no non-linear effects spoils the analysis. The low orders of detrending
seems to be inefficient and too high orders are influenced by numeric
instability but the main message is robust.

The analysis of the spatial properties of the noise within the electric
grids among the states does not differ so much. The properties of
the distribution function among the states are practically the same
but the differences are present for the autocorrelation function.
Few states possess the most frequent description by the derivatives
of the power load and the weak anti-persistent process. Few states
(Luxembourg, Latvia and North Ireland) follow the strong persistent
process of the electric power loads and  most  states follow description
by the derivatives and strong anti-persistent process. The wide width
of the multifractal spectrums is mostly scattered among Europe  except
for the cluster of  central European countries where the width of
the shuffled multifractal spectrum  can be understood as an effect
of the imbalance due to installations of the wind power plants in
north Germany. The problem should be also addressed more closely using
different datasets. Interesti

The wide width of the autocorrelation multifractal spectrum is scattered
on the map of Europe. Since it is present for Iceland we believe that
it shall be related to  the hidden connection with the structure
of the power production. We emphasize that more analysis of the wider
set of datasets is needed.

To put the paper in the context of the analysis of the time series
and particularly in the analysis electric electric grids, we exploited
widely used methodology and improved it with the additional tests
and detrending method. We performed deep analysis of stochastic properties
of the time series for European countries that goes beyond trends
and we address the properties of the noise and its effective description
by Langevin equation. 

Practically, in the long horizon, the study shall be useful for unbiased
description of the electric power grids using the stochastic processes
and the effective handling of the risk management. Based on the analysis
we focus on use of the Fractionally integrated autoregressive conditional
heteroscedastic processes (FIARCH), see \cite{Granger-Ding:1996,Rangarajan_Ding_Long-Range_correlations}
or Autoregressive fractionally integrated moving average process (ARFIMA)
\cite{key70,Hosking1981,Granger-Joyeux:1980} that are able to generate
the time series with the power-law correlations for prediction of
the electric power loads. Recently, generalization of both processes
is proposed in \cite{Podobnik-et-all:2005} for symmetric distribution
functions.  However, our previous study \cite{kracik-lavicka:2016}
revealed that the distribution function is asymmetric even more it
may not fit to any element from the set of the Lévy stable distributions,
see \cite{key55,key70,Podobnik-et-all:2005}.  It supports a hypothesis
that the poles of the Mellin transform, see \cite{Kilbas_Saigo_H_transforms},
of the distributions are not stationary.

\section*{Author contributions}

J.K. obtained and prepared the dataset. J.K and H.L. developed the
main ideas of the paper. H.L. prepared the tool for analysis for UNIX-like
system. Both authors also performed literature search, the analysis
of the time series and the visualization of the results. J.K and H.L
both contributed to the writing of the manuscript and the work was
performed under H.L. leadership. The work described in this paper
will be used in J.K.'s Ph.D. thesis.
\begin{acknowledgments}
The analysis exhibits personal view of authors of electroenergetics.
No authority or a company did influence the analysis. We acknowledge
fruitful discussions with P. Jizba, J. Lavi\v{c}ka, E. Lutz, T. Kiss,
G. Alber, E. Gil, R. Weron, M. Ausloos, L.Matsuoka and H.E. Stanley.
It was also supported by Czech Ministry of Education RVO68407700.
This thesis benefited from the European Union's Horizon 2020 Research
and Innovation Staff Exchange programme under the Marie Sklodowska-Curie
grant agreement No 681228.
\end{acknowledgments}

\begin{table*}
\begin{tabular}{|c|c|c|c|c|c|c|c|c|c|c|}
\hline 
 & \multicolumn{5}{c|}{Widths of multifractal spectrum} & \multicolumn{5}{c|}{Position of maxima}\tabularnewline
\hline 
Country & $\Delta\pi$ & $\Delta\pi^{shuf}$ & $\bigtriangleup\pi^{sur}$ & $\bigtriangleup\pi^{cor}$ & $\bigtriangleup\pi^{dist}$ & $\pi_{max}$ & $\pi_{max}^{shuf}$ & $\pi_{max}^{sur}$ & $\pi_{max}^{cor}$ & $\pi_{max}^{dist}$\tabularnewline
\hline 
\hline 
Austria & 0.370  & 0.175 & 0.362  & 0.216  & 0.102  & 1.266  & 0.494  & 1.232  & 0.772  & -0.018\tabularnewline
\hline 
Belgium & 0.383  & 0.215 & 0.373 & 0.199 & 0.099  & 1.123 & 0.499 & 1.110 & 0.624 & -0.029\tabularnewline
\hline 
Bosnia-Herzegovina & 0.363  & 0.190 & 0.310 & 0.184 & 0.095 & 1.137 & 0.493 & 1.116 & 0.644 & -0.008\tabularnewline
\hline 
Bulgaria & 0.378  & 0.199 & 0.335 & 0.192 & 0.064 & 1.121 & 0.500 & 1.120 & 0.621 & -0.010\tabularnewline
\hline 
Croatia & 0.374  & 0.199 & 0.304 & 0.176 & 0.071 & 1.169 & 0.497 & 1.154 & 0.672 & 0.010\tabularnewline
\hline 
Cyprus & 0.461  & 0.199 & 0.412 & 0.290 & 0.089 & 1.255 & 0.495 & 1.279 & 0.760 & -0.017\tabularnewline
\hline 
Czech Republic & 0.426  & 0.227 & 0.380 & 0.209 & 0.090 & 1.133 & 0.501 & 1.122 & 0.632 & -0.021\tabularnewline
\hline 
Denmark\footnote{Excludes oversea dominions.\label{fn:Excludes-oversea-dominions.}} & 0.487  & 0.208 & 0.359  & 0.300 & 0.155 & 1.240 & 0.495 & 1.217 & 0.745 & 0.022\tabularnewline
\hline 
Estonia & 0.559  & 0.226 & 0.487 & 0.383 & 0.185 & 1.113 & 0.498 & 1.138 & 0.605 & -0.093\tabularnewline
\hline 
Finland & 0.374  & 0.179 & 0.341 & 0.209 & 0.099 & 1.283 & 0.497 & 1.293 & 0.786 & -0.053\tabularnewline
\hline 
France\textsuperscript{\ref{fn:Excludes-oversea-dominions.}} & 0.371  & 0.210 & 0.367 & 0.169 & 0.057 & 1.227 & 0.495 & 1.223 & 0.724 & -0.037\tabularnewline
\hline 
Germany & 0.432  & 0.249 & 0.415 & 0.225 & 0.085 & 1.087 & 0.502 & 1.091 & 0.583 & -0.023\tabularnewline
\hline 
United Kingdom\footnote{Excluding North Ireland, Guersey, Jersey, Isle of Man and oversea
dominions, includes England, Wales and Scotland.} & 0.404  & 0.216 & 0.368 & 0.194 & 0.071 & 1.126 & 0.499 & 1.138 & 0.620 & -0.045\tabularnewline
\hline 
Greece & 0.408  & 0.196 & 0.341 & 0.215 & 0.092 & 1.195 & 0.492 & 1.177 & 0.703 & -0.001\tabularnewline
\hline 
Hungary & 0.458  & 0.196 & 0.352 & 0.286 & 0.148 & 1.158 & 0.493 & 1.126 & 0.665 & 0.033\tabularnewline
\hline 
Ireland & 0.386  & 0.219 & 0.333 & 0.186 & 0.085 & 1.060 & 0.494 & 1.061 & 0.565 & -0.002\tabularnewline
\hline 
Iceland & 0.570  & 0.201 & 0.434 & 0.417 & 0.214 & 1.090 & 0.496 & 1.088 & 0.597 & -0.030\tabularnewline
\hline 
Italy & 0.433  & 0.200 & 0.387 & 0.271 & 0.146 & 1.278 & 0.498 & 1.236 & 0.780 & 0.003\tabularnewline
\hline 
Latvia & 0.351  & 0.213 & 0.302 & 0.143 & 0.058 & 1.018 & 0.496 & 1.017 & 0.521 & 0.000\tabularnewline
\hline 
Lithuania & 0.457  & 0.202 & 0.387 & 0.275 & 0.138 & 1.257 & 0.497 & 1.228 & 0.760 & 0.014\tabularnewline
\hline 
Luxembourg & 0.356  & 0.186 & 0.269 & 0.178 & 0.092 & 0.951 & 0.494 & 0.973 & 0.460 & -0.010\tabularnewline
\hline 
Macedonia\footnote{FYROM Macedonia} & 0.387  & 0.174 & 0.308 & 0.219 & 0.086 & 1.283 & 0.489 & 1.265 & 0.794 & 0.028\tabularnewline
\hline 
Montenegro & 0.377  & 0.169 & 0.335 & 0.213 & 0.111 & 1.268 & 0.491 & 1.249 & 0.762 & -0.063\tabularnewline
\hline 
Netherlands\textsuperscript{\ref{fn:Excludes-oversea-dominions.}} & 0.498  & 0.182 & 0.390 & 0.341 & 0.171 & 1.183 & 0.495 & 1.160 & 0.688 & 0.035\tabularnewline
\hline 
Northern Ireland\footnote{Part of Great Britain with separated power grid system.} & 0.381  & 0.246 & 0.357 & 0.149 & 0.080 & 0.952 & 0.500 & 0.954 & 0.450 & -0.049\tabularnewline
\hline 
Norway & 0.369  & 0.214 & 0.339 & 0.166 & 0.075 & 1.231 & 0.494 & 1.257 & 0.736 & -0.060\tabularnewline
\hline 
Poland & 0.536  & 0.249 & 0.469 & 0.340 & 0.187 & 1.196 & 0.502 & 1.141 & 0.694 & -0.040\tabularnewline
\hline 
Portugal & 0.466  & 0.203 & 0.339 & 0.284 & 0.179 & 1.175 & 0.496 & 1.146 & 0.679 & 0.046\tabularnewline
\hline 
Romania & 0.409  & 0.191 & 0.361 & 0.230 & 0.108 & 1.183 & 0.495 & 1.169 & 0.688 & -0.032\tabularnewline
\hline 
Serbia & 0.396  & 0.175 & 0.315 & 0.229 & 0.107 & 1.204 & 0.494 & 1.197 & 0.709 & -0.013\tabularnewline
\hline 
Slovakia & 0.385  & 0.223 & 0.364 & 0.180 & 0.092 & 1.183 & 0.500 & 1.167 & 0.683 & -0.024\tabularnewline
\hline 
Slovenia & 0.388  & 0.198 & 0.299 & 0.200 & 0.107 & 1.157 & 0.495 & 1.131 & 0.662 & 0.026\tabularnewline
\hline 
Spain\textsuperscript{\ref{fn:Excludes-oversea-dominions.}} & 0.574 & 0.258 & 0.396 & 0.354 & 0.209 & 1.169 & 0.504 & 1.131 & 0.665 & 0.038\tabularnewline
\hline 
Sweden & 0.402 & 0.180 & 0.325 & 0.224  & 0.079 & 1.300  & 0.494 & 1.331 & 0.807 & -0.027\tabularnewline
\hline 
Switzerland & 0.390 & 0.175 & 0.310 & 0.226 & 0.114 & 1.214 & 0.500 & 1.204 & 0.714 & 0.014\tabularnewline
\hline 
\end{tabular}

\caption{Collection of results of the MFDFA analysis of the order $4$.}
\label{multifractal_width_table}
\end{table*}

\bibliographystyle{ieeetr}
\bibliography{bibliography}

\end{document}